\documentclass[english]{revtex4-1}
\usepackage[T1]{fontenc}
\usepackage[latin9]{inputenc}
\usepackage{amsmath}
\usepackage{amssymb}
\usepackage{graphicx}

\makeatletter
\usepackage{babel}

\makeatother

\usepackage{babel}
\begin{document}

\title{Electromagnetic field expectations as measures of photon localization}

\author{Scott E. Hoffmann}

\address{School of Mathematics and Physics,~~\\
 The University of Queensland,~~\\
 Brisbane, QLD 4072~~\\
 Australia}
\email{scott.hoffmann@uqconnect.edu.au}

\selectlanguage{english}%
\begin{abstract}
The questions of whether a photon can be localized in an arbitrarily
small volume and what is the allowable strength of that localization
(the decrease with distance of the functional form) are questions
of current interest. We propose a measure of localization for the
single photon that is the expectation values of the electromagnetic
field strength components in a coherent wavepacket state of mean photon
number unity. As such, we deal with real quantities that have a physical
meaning rather than complex amplitudes. It is seen that the real parts
of complex amplitudes proposed previously as measures of localization
are equal to our field expectations. With this measure, we examine
two test states. The first has a well-resolved momentum. The field
expectations show Gaussian (quadratic exponential) localization in
all directions, although the localization length scale is much larger
than the mean wavelength. For the other test state, with a spherically
symmetric momentum distribution, we find almost exponential localization
in all directions at $t=0.$ The profiles are scale invariant, so
choosing the momentum width very large would make the localization
length arbitrarily small. We conclude that there is no lower bound
on the localization length scale of a photon as determined by this
measure.
\end{abstract}
\maketitle

\section{Introduction}

A photon cannot be localized at a point, according to the three criteria
of Newton and Wigner \cite{Newton1949}. The reason is not its masslessness,
as a hypothetic particle of zero mass and zero helicity could be localized
at a point. The reason, instead, is due to the limited helicity spectrum
of the photon, as pointed out by Wightman \cite{Wightman1962}. If
there were a photon of zero helicity that entered into superpositions
with the other two helicities, $\lambda=\pm1,$ point localized states
could be constructed, with an extra, spin-like quantum number carrying
an irreducible (spin 1) representation of rotations.

Experiments, though, suggest that a photon can be localized, not at
a point, but with a high probability in a small volume of space. A
converging lens will focus a laser to a spot size of order the wavelength.
A photon state vector collapses to a well-localized state vector when
it darkens a crystal of photographic emulsion. Thus we need measures
of localization to quantify this observed probabilistic localization.

To restate the limitations, there can be no point localized state
vectors of the form $|\,t,\boldsymbol{r},\mu\,\rangle$ for $\mu=-s,\dots,s$
satisfying
\begin{equation}
\langle\,t,\boldsymbol{r}_{1},\mu_{1}\,|\,t,\boldsymbol{r}_{2},\mu_{2}\,\rangle=\delta_{\mu_{1}\mu_{2}}\delta^{3}(\boldsymbol{r}_{1}-\boldsymbol{r}_{2})\quad\mathrm{and}\quad U(R)\,|\,t,\boldsymbol{0},\mu\,\rangle=\sum_{\mu^{\prime}=-s}^{s}|\,t,\boldsymbol{0},\mu^{\prime}\,\rangle\mathcal{D}_{\mu^{\prime}\mu}^{(s)}(R),\label{eq:1.1}
\end{equation}
with smooth boost behaviour, for any rotation representation label
$s,$ integral. There can be no position probability amplitudes of
the form
\begin{equation}
\psi_{\mu}(x)=\langle\,x,\mu\,|\,\psi\,\rangle,\label{eq:1.2}
\end{equation}
with $x^{\nu}=(t,\boldsymbol{r})^{\nu}.$ There can be no position
operator (at $t=0$) of the form
\begin{equation}
\hat{\boldsymbol{x}}=\int d^{3}r\sum_{\mu=-s}^{s}|\,0,\boldsymbol{r},\mu\,\rangle\boldsymbol{r}\langle\,0,\boldsymbol{r},\mu\,|.\label{eq:1.3}
\end{equation}

Within these limitations, we do not look for an amplitude whose modulus
squared (perhaps summed over an extra index) is to be a measure of
localization. Instead, we consider expectation values of the six components
of the gauge-invariant electromagnetic field strength operators. These
are well-defined for a system of free photons (Eq. (\ref{eq:6.1}))
and are Hermitian observables corresponding to physical measurements.
It might be experimentally difficult to measure field strengths for
a single photon, but this will give a consistent theoretical characterization.
The only problem is that the field strengths, depending on the creation
and annihilation operators, change the photon number by $\pm1$ and
so have vanishing expectation values in a state of definite photon
number. We avoid this problem by calculating expectation values in
a coherent wavepacket state, a superposition of all photon number
states, including the vacuum, with a mean photon number of one. While
it is possible to make weak coherent states experimentally \cite{Izumi2018},
the construction of this experiment would be challenging. But again,
we are only looking for a consistent theoretical characterization.
This measure of localization is what quantum mechanics predicts for
the result of repeated measurements of Hermitian observables.

Other authors have constructed wavefunctions for the photon \cite{Sipe1995,Landau1930,Bialynicki-Birula1994,Silberstein1907a,Silberstein1907b}.
None of these are of the form of Eq. (\ref{eq:1.2}). Photon position
operators have been proposed \cite{Hawton1999,Pryce1948}, none of
which can be brought to the form of Eq. (\ref{eq:1.3}). Other authors
have investigated limits to the degree of localization of the photon,
meaning the rate of falloff of the measure of localization for large
distances from the point of average localization \cite{Saari2012,Saari2005,Bialynicki-Birula1998}.
In this paper, we construct a photon state vector that is seen to
have strong, Gaussian, localization along the $x,y$ and $z$ axes.
Since a boost will Lorentz contract any spatial feature along the
direction of the boost, we see that a photon must be localizable in
an arbitrarily small length scale in that direction. That strong localization
is possible in all directions is not surprising from the Heisenberg
uncertainty principle.

The organization of this paper is as follows. We start, in Section
\ref{sec:Poincar=0000E9-representations-for}, by reviewing the Poincaré
transformation properties of the momentum-helicity basis vectors,
$|\,k,\lambda\,\rangle,$ and the momentum-helicity probability amplitudes,
$\Psi_{\lambda}(k),$ for $k^{2}=0,k^{0}=\omega=|k|$ and $\lambda=\pm1.$
Any proposed amplitude describing the photon must be a linear function
of the $\Psi_{\lambda}(k).$ We provide a physical interpretation
of the little group elements and expressions for the Wigner phases
for any rotation and any boost.

In Section \ref{sec:Measurement-of-localization}, we consider the
measures of localization that are the expectation values of the gauge-invariant
electromagnetic field strength components in a coherent state. As
such, they are not linear in the $\Psi_{\lambda}(k).$ We consider
a test state with a spherically symmetric momentum distribution, designed
for localization in a small volume for large momentum spread, $\sigma_{k}.$
We plot field component expectations along the $x,y$ and $z$ axes.
We find localization with an almost exponential asymptotic form, in
contrast to the Gaussian localization for our previous example.

In Section \ref{sec:Comparison-with-other}, we consider the work
of other authors on proposed wavefunctions for the photon and on the
degree of localizability of the photon.

Conclusions follow in Section \ref{sec:Conclusions}.

Throughout this paper, we use Heaviside-Lorentz units, in which $\hbar=c=\epsilon_{0}=\mu_{0}=1$.

\section{\label{sec:Poincar=0000E9-representations-for}Poincaré representations
for the photon}

The unitary, irreducible representations of the Poincaré group for
the photon are carried by the improper basis vectors $|\,k,\lambda\,\rangle$
for $\lambda=\pm1.$ These are eigenvectors of four-momentum with
eigenvalue components $k^{\mu}=(\omega,\boldsymbol{k})^{\mu}$ satisfying
$k^{2}=0.$ The positive energies are $k^{0}=\omega(\boldsymbol{k})=|\boldsymbol{k}|.$
By giving them the covariant normalization
\begin{equation}
\langle\,k_{1},\lambda_{1}\,|\,k_{2},\lambda_{2}\,\rangle=\delta_{\lambda_{1}\lambda_{2}}\omega_{1}\delta^{3}(\boldsymbol{k}_{1}-\boldsymbol{k}_{2}),\label{eq:2.1}
\end{equation}
the following transformation laws take their simplest forms. The four-momentum
carries a one-dimensional representation of spacetime translations,
\begin{equation}
U(T(a))\,|\,k,\lambda\,\rangle=|\,k,\lambda\,\rangle\,e^{ik\cdot a}.\label{eq:2.2}
\end{equation}
The helicity, $\lambda,$ a Poincaré invariant, carries a one-dimensional
representation of rotations about the three-momentum direction,
\begin{equation}
U(R(\Omega\,\hat{\boldsymbol{k}}))\,|\,k,\lambda\,\rangle=|\,k,\lambda\,\rangle\,e^{-i\lambda\Omega}.\label{eq:2.3}
\end{equation}
That the photon has two and only two polarizations, with one or minus
one units of angular momentum in the momentum direction, is an experimental
result \cite{ParticleDataGroup2018}. The positive and negative helicities
correspond to left- and right-circular polarization, respectively.
Linear superpositions of these can be constructed to describe linear
polarizations mutually orthogonal and orthogonal to the momentum direction.

To find the effects of general Lorentz transformations on the basis
vectors requires construction of a vector, rather than ray, representation
\cite{Halpern1968}. In this, the phase of every basis vector is uniquely
defined in terms of that of, in this case, two reference state vectors
with helicities $\lambda=\pm1.$ To construct this representation
requires knowledge of the little group for a particle momentum, which
is the subgroup of the Lorentz group that leaves that momentum unchanged.
For massive particles there is always a rest frame, and the little
group is the group of rotations in the rest frame. We are familiar
with the unitary, irreducible representations of rotations, with the
consequence that a massive particle can have a spin that can take
any value $s=0,\frac{1}{2},1,\frac{3}{2},\dots.$ There is no rest
frame for a massless particle, so we choose as the reference state
vectors $|\,\kappa,\lambda\,\rangle$ for $\lambda=\pm1,$ with a
particular choice, $\kappa,$ of the energy and a four-momentum $k_{R}^{\mu}=(\kappa,0,0,\kappa)^{\mu}.$

Clearly a rotation about the $z$ direction leaves this four-momentum
unchanged. The set of all such rotations is then a subgroup of the
little group for this massless momentum. The helicity appearing in
the one-dimensional representations could, in principle, take any
integral or half-integral value.

Only for massless particles, there are boosts that change the direction
of the momentum but leave the energy unchanged. We call these isoenergetic
boosts. If such a transformation is followed by a rotation that takes
the momentum back to the $z$ direction, we have an element of the
other subgroup of the little group. We call this Lorentz transformation
an IBR, an isoenergetic boost followed by a rotation.

We find that if the isoenergetic boost direction has spherical polar
angles $(\theta_{B},\varphi_{B})$, then the isoenergetic boost velocity
must be
\begin{equation}
\boldsymbol{\beta}(\theta_{B},\varphi_{B})=-\frac{2\cos\theta_{B}}{1+\cos^{2}\theta_{B}}\,\hat{\boldsymbol{u}}(\theta_{B},\varphi_{B}),\label{eq:2.4}
\end{equation}
where $\hat{\boldsymbol{u}}(\theta_{B},\varphi_{B})$ is a unit vector
with those spherical polar angles. Note that the speed as written
is less than unity on $0<\theta_{B}<\pi$ and is negative on $0<\theta_{B}<\pi/2.$
The final polar angle of the boosted momentum is found to be
\begin{equation}
\psi(\theta_{B})=2\theta_{B}-\pi,\label{eq:2.5}
\end{equation}
with $-\pi<\psi(\theta_{B})<\pi.$

If we define
\begin{equation}
\boldsymbol{\alpha}\equiv-2\cot\theta_{B}\,(\cos\varphi_{B}\,\hat{\boldsymbol{x}}+\sin\varphi_{B}\,\hat{\boldsymbol{y}}),\label{eq:2.6}
\end{equation}
and $\hat{\boldsymbol{u}}_{1}=\cos\varphi_{B}\,\hat{\boldsymbol{x}}+\sin\varphi_{B}\,\hat{\boldsymbol{y}}$,
we find that the IBR has the Lorentz transformation matrix
\begin{equation}
\mathcal{L}_{\phantom{\mu}\nu}^{\mu}(\boldsymbol{\alpha})=[R(-\psi(\theta_{B})\hat{\boldsymbol{u}}_{2})\Lambda(\boldsymbol{\beta}(\theta_{B},\varphi_{B}))]_{\phantom{\mu}\nu}^{\mu}=\begin{pmatrix}1+\frac{1}{2}\boldsymbol{\alpha}^{2} & \alpha_{x} & \alpha_{y} & -\frac{1}{2}\boldsymbol{\alpha}^{2}\\
\alpha_{x} & 1 & 0 & -\alpha_{x}\\
\alpha_{y} & 0 & 1 & -\alpha_{y}\\
\frac{1}{2}\boldsymbol{\alpha}^{2} & \alpha_{x} & \alpha_{y} & 1-\frac{1}{2}\boldsymbol{\alpha}^{2}
\end{pmatrix}_{\phantom{\mu}\nu}^{\mu},\label{eq:2.7}
\end{equation}
where $\hat{\boldsymbol{u}}_{2}=\hat{\boldsymbol{u}}_{1}\times\hat{\boldsymbol{z}}.$

From this form, we can derive the group multiplication laws
\begin{align}
\mathcal{L}(\boldsymbol{\alpha}_{1})\mathcal{L}(\boldsymbol{\alpha}_{2}) & =\mathcal{L}(\boldsymbol{\alpha}_{1}+\boldsymbol{\alpha}_{2}),\nonumber \\
R_{z}(\gamma)\mathcal{L}(\boldsymbol{\alpha})R_{z}^{-1}(\gamma) & =\mathcal{L}(R_{z}(\gamma)\boldsymbol{\alpha}).\label{eq:2.8}
\end{align}
So we see that the little group is isomorphic to the group of translations
and rotations in a plane, the Euclidean group in two dimensions. It
can be shown that the two commuting generators of the IBRs, with respect
to $\alpha_{x}$ and $\alpha_{y},$ are
\begin{eqnarray}
L_{x} & = & K_{x}-J_{y},\nonumber \\
L_{y} & = & K_{y}+J_{x},\label{eq:2.9}
\end{eqnarray}
in terms of the boost generators, $\boldsymbol{K},$ and the angular
momenta, $\boldsymbol{J}.$ The generator of the $z$ rotations with
respect to the rotation angle is $J_{z}.$

We invoke the assumption, motivated by experimental results and theoretical
considerations, that the state of a photon is completely characterized
by its momentum and helicity, with no reference to any other quantum
numbers that might carry a representation of an additional internal
symmetry group. Thus the isoenergetic boosts followed by rotations
must all be represented by unity acting on a physical photon state.

To construct momentum/helicity eigenvectors of general massless four-momentum,
$k^{\mu}=(\omega,\boldsymbol{k})^{\mu},$ we first boost the reference
state by
\begin{equation}
\boldsymbol{\beta}(\omega,\kappa)=\frac{\omega^{2}-\kappa^{2}}{\omega^{2}+\kappa^{2}}\hat{\boldsymbol{z}}\label{eq:2.10}
\end{equation}
to produce energy $\omega,$ with the transformation denoted $\Lambda_{z}(\omega,\kappa).$
This transformation commutes with rotations about the $z$ axis, so
it leaves helicity unchanged. Then we rotate into the direction $\hat{\boldsymbol{k}}=(\theta,\varphi)$
using the standard rotation
\begin{equation}
R_{0}[\hat{\boldsymbol{k}}]=R_{z}(\varphi)R_{y}(\theta)R_{z}(-\varphi).\label{eq:2.11}
\end{equation}
This is
\begin{equation}
|\,k,\lambda\,\rangle=U(R_{0}[\hat{\boldsymbol{k}}])U(\Lambda_{z}(\omega_{0},\kappa))\,|\,\kappa,\lambda\,\rangle=U(L(k,\kappa))\,|\,\kappa,\lambda\,\rangle.\label{eq:2.12}
\end{equation}

Now we can find the transformation properties of these basis vectors.
For a rotation, we find
\begin{align}
U(R)\,|\,k,\lambda\,\rangle & =|\,Rk,\lambda\,\rangle\,e^{-i\lambda w(R,k)},\label{eq:2.13}
\end{align}
where $w(R,k)$ is a Wigner rotation angle. It can be calculated explicity
by using the $j=\frac{1}{2}$ rotation representation to multiply
the little group element
\begin{equation}
R_{z}(w(R,k))=R_{0}^{-1}[R\hat{\boldsymbol{k}}]\,R\,R_{0}[\hat{\boldsymbol{k}}].\label{eq:2.14}
\end{equation}
We find
\begin{equation}
e^{-iw(R,k)}=\frac{\mathcal{R}_{++}\cos\frac{\theta}{2}+\mathcal{R}_{+-}\sin\frac{\theta}{2}\,e^{+i\varphi}}{\mathcal{R}_{++}^{*}\cos\frac{\theta}{2}+\mathcal{R}_{+-}^{*}\sin\frac{\theta}{2}\,e^{-i\varphi}},\label{eq:2.15}
\end{equation}
where
\begin{align}
\begin{pmatrix}\mathcal{R}_{++} & \mathcal{R}_{+-}\\
\mathcal{R}_{-+} & \mathcal{R}_{--}
\end{pmatrix}_{m_{1}m_{2}} & =\mathcal{D}_{m_{1}m_{2}}^{(\frac{1}{2})}(R)\nonumber \\
 & =\begin{pmatrix}\cos\frac{\Omega}{2}-i\sin\frac{\Omega}{2}\,\hat{\Omega}_{z} & -i(\hat{\Omega}_{x}-i\hat{\Omega}_{y})\sin\frac{\Omega}{2}\\
-i(\hat{\Omega}_{x}+i\hat{\Omega}_{y})\sin\frac{\Omega}{2} & \cos\frac{\Omega}{2}+i\sin\frac{\Omega}{2}\,\hat{\Omega}_{z}
\end{pmatrix}_{m_{1}m_{2}}\nonumber \\
 & =\begin{pmatrix}\cos\frac{\beta}{2}\,e^{-i(\alpha+\gamma)/2} & -\sin\frac{\beta}{2}\,e^{-i(\alpha-\gamma)/2}\\
\sin\frac{\beta}{2}\,e^{+i(\alpha-\gamma)/2} & \cos\frac{\beta}{2}\,e^{+i(\alpha+\gamma)/2}
\end{pmatrix}_{m_{1}m_{2}}\label{eq:2.16}
\end{align}
for a rotation parameterized by angle $\Omega$ about axis $\hat{\boldsymbol{\Omega}}$
or parameterized by Euler angles $\alpha,\beta,\gamma$ as $R=R_{z}(\alpha)R_{y}(\beta)R_{z}(\gamma).$
We find agreement with Caban \textit{et al.} \cite{Caban2003} (their
Equation (29)) after noting that, with their conventions (the same
as our conventions), they are calculating $\exp(-iw(R,k)),$ not $\exp(+iw(R,k))$
as written.

For a boost, we use
\begin{equation}
U(\Lambda)\,|\,k,\lambda\,\rangle=U(L(\Lambda k,\kappa))\{U^{\dagger}(\Lambda_{z}(\omega_{0}^{\prime},\kappa))U^{\dagger}(R_{0}[\Lambda k])U(\Lambda)U(R_{0}[k])U(\Lambda_{z}(\omega_{0},\kappa))\}\,|\,\kappa,\lambda\,\rangle.\label{eq:2.17}
\end{equation}
(For notational convenience, we have written standard rotations, $R_{0},$
depending on the four-vectors $k$ and $k^{\prime}=\Lambda k.$ By
that we mean that they depend on the unit vectors $\hat{\boldsymbol{k}}$
and $\hat{\boldsymbol{k}}^{\prime},$ respectively.) The transformation
in braces represents a little group element of the form $U(R_{z}(w(\Lambda,k)))U(\mathcal{L}(\boldsymbol{\alpha})).$
There is no need to calculate the parameter $\boldsymbol{\alpha}.$
We multiply matrices in the $(\frac{1}{2},0)$ nonunitary, finite
dimensional, representation of the Lorentz group \cite{Halpern1968}.
We parametrize the boost, $\Lambda,$ by the rapidity, $\boldsymbol{\zeta},$
related to the boost velocity by $\boldsymbol{\beta}=\tanh\zeta\,\hat{\boldsymbol{\zeta}.}$
In that representation an IBR is represented by
\begin{equation}
\mathrm{D}(\mathcal{L}(\boldsymbol{\alpha}))=\begin{pmatrix}1 & 0\\
-(\alpha_{x}+i\alpha_{y}) & 1
\end{pmatrix}.\label{eq:2.18}
\end{equation}
So we choose to apply the matrices to the spinor
\begin{equation}
v=\begin{pmatrix}0\\
1
\end{pmatrix},\label{eq:2.19}
\end{equation}
which is left unchanged by this transformation. We find

\begin{equation}
e^{-iw(\Lambda[\boldsymbol{\zeta}],k)}=\frac{(1+\tanh\frac{\zeta}{2}\,\hat{\zeta}_{z})\cos\frac{\theta}{2}+\tanh\frac{\zeta}{2}\,(\hat{\zeta}_{x}-i\hat{\zeta}_{y})\,\sin\frac{\theta}{2}\,e^{+i\varphi}}{(1+\tanh\frac{\zeta}{2}\,\hat{\zeta}_{z})\cos\frac{\theta}{2}+\tanh\frac{\zeta}{2}\,(\hat{\zeta}_{x}+i\hat{\zeta}_{y})\,\sin\frac{\theta}{2}\,e^{-i\varphi}}.\label{eq:2.20}
\end{equation}
Caban \textit{et al.} \cite{Caban2003} constructed their boost as
a boost in the $z$ direction followed by a rotation, so direct comparison
with our result is not possible. This expression reduces to unity
for $\hat{\boldsymbol{\zeta}}=\hat{\boldsymbol{k}},$ as expected.

A general, normalized, state vector can be written
\begin{equation}
|\,\psi\,\rangle=\int\frac{d^{3}k}{\sqrt{\omega}}\sum_{\lambda=\pm1}|\,k,\lambda\,\rangle\Psi_{\lambda}(k).\label{eq:2.21}
\end{equation}
The normalization condition is
\begin{equation}
\int d^{3}k\sum_{\lambda=\pm1}|\Psi_{\lambda}(k)|^{2}=1.\label{eq:2.22}
\end{equation}
The average momentum formula is
\begin{equation}
\langle\,\psi\,|\,P^{\mu}\,|\,\psi\,\rangle=\int d^{3}k\sum_{\lambda=\pm1}|\Psi_{\lambda}(k)|^{2}\,k^{\mu},\label{eq:2.23}
\end{equation}
where the Hamiltonian ($H=P^{0}$) and linear momentum operator are
the components of
\begin{equation}
P^{\mu}=\int\frac{d^{3}k}{\omega}\,|\,k,\lambda\,\rangle\,k^{\mu}\,\langle\,k,\lambda\,|.\label{eq:2.24}
\end{equation}
The average helicity formula is
\begin{equation}
\langle\,\psi\,|\,\hat{\lambda}\,|\,\psi\,\rangle=\int d^{3}k\sum_{\lambda=\pm1}|\Psi_{\lambda}(k)|^{2}\,\lambda,\label{eq:2.25}
\end{equation}
(where $\hat{\lambda}\,|\,k,\lambda\,\rangle=\lambda\,|\,k,\lambda\,\rangle$).
Together, these confirm the interpretation of $\Psi_{\lambda}(k)$
as a momentum-helicity probability amplitude.

Then for the action of a unitary or antiunitary transformation on
$|\,\psi\,\rangle$ defined by
\begin{equation}
U/A\,|\,\psi\,\rangle=\int\frac{d^{3}k}{\sqrt{\omega}}\sum_{\lambda=\pm1}|\,k,\lambda\,\rangle\Psi_{\lambda}^{\prime}(k),\label{eq:2.26}
\end{equation}
we find the transformation properties ($\hat{\boldsymbol{k}}=(\theta,\varphi)$)
\begin{eqnarray}
\mathrm{Spacetime\ translations:}\quad\Psi_{\lambda}^{\prime}(k) & = & e^{+ik\cdot a}\,\Psi_{\lambda}(k),\nonumber \\
\mathrm{Rotations:}\quad\Psi_{\lambda}^{\prime}(k) & = & e^{-i\lambda w(R,R^{-1}k)}\,\Psi_{\lambda}(R^{-1}k),\nonumber \\
\mathrm{Boosts\ by\ velocity\ \boldsymbol{\beta}:}\quad\Psi_{\lambda}^{\prime}(k) & = & \sqrt{\gamma(1-\boldsymbol{\beta}\cdot\hat{\boldsymbol{k}})}\,e^{-i\lambda w(\Lambda,\Lambda^{-1}k)}\,\Psi_{\lambda}(\Lambda^{-1}k),\nonumber \\
\mathrm{Space\ inversion}:\quad\Psi_{\lambda}^{\prime}(\omega_{0},\boldsymbol{k}) & = & \eta\,e^{+i2\lambda\varphi}\,\Psi_{-\lambda}(\omega_{0},-\boldsymbol{k}),\nonumber \\
\mathrm{Time\ reversal}:\quad\Psi^{\prime}(\omega_{0},\boldsymbol{k}) & = & e^{-i2\lambda\varphi}\,\Psi_{\lambda}^{*}(\omega_{0},-\boldsymbol{k}),\label{eq:2.27}
\end{eqnarray}
where $\eta=-1$ is the intrinsic parity of the photon \cite{ParticleDataGroup2018}.
Note that only space inversion changes the helicity.

\section{\label{sec:Measurement-of-localization}Measurement of localization
with electromagnetic field strengths}

The measure of localization that we consider here is the expectations
of the electromagnetic field strength operators. In the space of an
arbitrary number of photons, these are \cite{Itzykson1980}
\begin{equation}
F^{\mu\nu}(x)=\frac{1}{\sqrt{16\pi^{3}}}\int\frac{d^{3}k}{\omega_{0}}\sum_{\lambda=\pm1}(k^{\mu}\epsilon^{\nu}(k,\lambda)-k^{\nu}\epsilon^{\mu}(k,\lambda))\,a(k,\lambda)\,e^{-ik\cdot x}+(\dagger),\label{eq:6.1}
\end{equation}
where $(\dagger)$ is the Hermitian conjugate of the first term. The
components, in terms of the electric fields, $\boldsymbol{E}(x),$
and the magnetic fields, $\boldsymbol{B}(x),$ are
\begin{equation}
F^{\mu\nu}=\begin{pmatrix}0 &  & -E_{j}\\
\\
+E_{i} &  & -\epsilon_{ijk}B_{k}\\
\\
\end{pmatrix}.\label{eq:6.2}
\end{equation}

The creation and annihilation operators have the commutators
\begin{equation}
[a(k_{1},\lambda_{1}),a^{\dagger}(k_{2},\lambda_{2})]=\delta_{\lambda_{1}\lambda_{2}}\,\omega_{1}\delta^{3}(\boldsymbol{k}_{1}-\boldsymbol{k}_{2}),\quad[a(k_{1},\lambda_{1}),a(k_{2},\lambda_{2})]=[a^{\dagger}(k_{1},\lambda_{1}),a^{\dagger}(k_{2},\lambda_{2})]=0.\label{eq:6.3}
\end{equation}

We comment briefly on the construction of the polarization vectors,
$\epsilon^{\mu}(k,\lambda),$ as it points out the connection between
gauge dependence and the little group. We would like these polarization
vectors to transform as four-vectors, but we find that this is not
possible. Their construction follows very closely the construction
of the set of state vectors in Section \ref{sec:Poincar=0000E9-representations-for},
by defining reference vectors, then boosting in the $z$ direction
(which does not change transverse components) and finally using the
standard rotation to define the vectors for arbitrary momentum direction
$\hat{\boldsymbol{k}}.$

The reference vectors are chosen with a vanishing zero component and
only transverse spatial components
\begin{equation}
\epsilon^{\mu}(\kappa,+1)=-\frac{1}{\sqrt{2}}(0,1,i,0)^{\mu}\quad\mathrm{and}\quad\epsilon^{\mu}(\kappa,-1)=+\frac{1}{\sqrt{2}}(0,1,-i,0)^{\mu},\label{eq:6.7}
\end{equation}
satisfying the Lorentz condition
\begin{equation}
k_{R}\cdot\epsilon(\kappa,\lambda)=0,\label{eq:6.8}
\end{equation}
with $k_{R}^{\mu}=(\kappa,0,0,\kappa)^{\mu}.$

Then for general momentum, the vectors are
\begin{equation}
\epsilon^{\mu}(k,\lambda)=L(k,\kappa)_{\phantom{\mu}\nu}^{\mu}\epsilon^{\nu}(\kappa,\lambda)=R_{0}[\hat{\boldsymbol{k}}]_{\phantom{\mu}\nu}^{\mu}\epsilon^{\nu}(\kappa,\lambda).\label{eq:6.9}
\end{equation}
Note that the Lorentz condition,
\begin{equation}
k\cdot\epsilon(k,\lambda)=0,\label{eq:6.10}
\end{equation}
holds for all momenta.

Then the action of a general rotation or boost on these vectors is
\begin{equation}
R_{\phantom{\mu}\nu}^{\mu}\epsilon^{\nu}(k,\lambda)=\epsilon^{\mu}(Rk,\lambda)\,e^{-i\lambda w(R,k)}\label{eq:6.11}
\end{equation}
and
\begin{equation}
\Lambda_{\phantom{\mu}\nu}^{\mu}\epsilon^{\nu}(k,\lambda)=\epsilon^{\mu}(\Lambda k,\lambda)\,e^{-i\lambda w(\Lambda,k)}+\boldsymbol{\alpha}\cdot\boldsymbol{\epsilon}(\kappa,\lambda)\frac{(\Lambda k)^{\mu}}{\kappa},\label{eq:6.12}
\end{equation}
for a parameter $\boldsymbol{\alpha}$ that we do not calculate. In
particular,
\begin{equation}
R(\Omega\,\hat{\boldsymbol{k}})_{\phantom{\mu}\nu}^{\mu}\epsilon^{\nu}(k,\lambda)=\epsilon^{\nu}(k,\lambda)\,e^{-i\lambda\Omega},\label{eq:6.13}
\end{equation}
as appropriate for the polarization vectors of helicity eigenvectors.

We see from the boost result that the polarization vectors are not
invariant under the IBR transformations and so do not transform as
four-vectors. Instead, the transformation involves the addition of
a part proportional to the four-momentum, which is a gauge transformation.
This result has been obtained previously by other authors \cite{Han1981}.

However, the tensor
\begin{equation}
T^{\mu\nu}(k,\lambda)=k^{\mu}\epsilon^{\nu}(k,\lambda)-k^{\nu}\epsilon^{\mu}(k,\lambda)\label{eq:6.14}
\end{equation}
is gauge invariant and thus transforms as a rank 2 tensor. This leads
to the correct transformation properties of the electromagnetic field
strengths.

They transform locally as the components of a rank 2 antisymmetric
tensor and translate locally. They satisfy the free Maxwell equations.
They are normalized to
\begin{equation}
\int d^{3}x\,:(\frac{1}{2}(\boldsymbol{E}^{2}+\boldsymbol{B}^{2}),\frac{1}{2}(\boldsymbol{E}\times\boldsymbol{B}-\boldsymbol{B}\times\boldsymbol{E}))^{\mu}:=\int\frac{d^{3}k}{\omega}\sum_{\lambda=\pm1}a^{\dagger}(k,\lambda)a(k,\lambda)\,k^{\mu},\label{eq:6.15}
\end{equation}
the total four-momentum operator in the space of an arbitrary number
of photons. The colons indicate normal ordering, where the creation
operators are placed to the left of the annihilation operators, and
is necessary to remove an infinite contribution.

Because the creation and annihilation operators change the photon
number by $\pm1,$ respectively, the electromagnetic field strengths
have vanishing expectation values in a state of a definite number
of photons. Instead we calculate nonvanishing expectations of the
field components in a coherent wavepacket state, which is a superposition
of all number states \cite{Glauber1963a,Glauber1963b}, with a mean
photon number of unity. Admittedly, a single photon state is not a
coherent state. We argue that the coherent wavepacket state we construct
is the closest we can come to capturing the essential physics of the
single photon. Sources of weak coherent states (from faint lasers)
are widely used in quantum measurement and technology \cite{Izumi2018},
and typically have an average photon number less than one \cite{Izumi2018}.
The expectations we calculate are the average values that would be
observed if field strength measurements were performed on a weak coherent
state with unit average photon number.

We use the wavepacket annihilation operator
\begin{equation}
a(\psi)=\int\frac{d^{3}k}{\sqrt{\omega}}\sum_{\lambda=\pm1}\Psi_{\lambda}^{*}(k)\,a(k,\lambda),\label{eq:6.23}
\end{equation}
satisfying
\begin{equation}
[a(\psi),a^{\dagger}(\psi)]=1,\quad[a(\psi),a(\psi)]=[a^{\dagger}(\psi),a^{\dagger}(\psi)]=0.\label{eq:6.24}
\end{equation}
Then a coherent wavepacket state with arbitrary mean photon number
(related to $z$) is required to satisfy (the Barut-Girardello definition
\cite{Barut1971})
\begin{equation}
a(\psi)\,|\,z,\psi\,\rangle=z\,|\,z,\psi\,\rangle.\label{eq:6.25}
\end{equation}
The solution is well known \cite{Glauber1963a,Glauber1963b}
\begin{equation}
|\,z,\psi\,\rangle=e^{-|z|^{2}/2}\sum_{n=0}^{\infty}\frac{(z\,a^{\dagger}(\psi))^{n}}{n!}\,|\,0\,\rangle.\label{eq:6.26}
\end{equation}
The mean photon number is found to be $\langle\,z,\psi\,|\,N\,|\,z,\psi\,\rangle=|z|^{2},$
so we choose $z=1$ to make a coherent wavepacket state that mimicks
a single photon state with that momentum-helicity wavefunction.

With this choice, we find the field expectation values become
\begin{equation}
\overline{F}^{\mu\nu}(x)=\langle\,1,\psi\,|\,F^{\mu\nu}(x)\,|\,1,\psi\,\rangle=\frac{1}{\sqrt{16\pi^{3}}}\int\frac{d^{3}k}{\omega}\sum_{\lambda=\pm1}(k^{\mu}\epsilon^{\nu}(k,\lambda)-k^{\nu}\epsilon^{\mu}(k,\lambda))\,\sqrt{\omega}\,\Psi_{\lambda}(k)\,e^{-ik\cdot x}+\mathrm{c}.\mathrm{c}.,\label{eq:6.26.a}
\end{equation}
where c.c. indicates the complex conjugate of the first term. Once
these are calculated, we can form what, classically, is an energy
density
\begin{equation}
\mathcal{E}(x)=\frac{1}{2}(\overline{\boldsymbol{E}}(x)^{2}+\overline{\boldsymbol{B}}(x)^{2}).\label{eq:6.26.1}
\end{equation}

This is normalized to
\begin{equation}
\int d^{3}x\,\mathcal{E}(x)=\langle\,1,\psi\,|\,H\,|\,1,\psi\,\rangle.\label{eq:6.21.2}
\end{equation}

For a state of well-resolved momentum and a definite helicity, we
choose the probability amplitudes to be
\begin{equation}
\Psi_{\lambda}(k)=\delta_{\lambda1}\,\frac{e^{-|\boldsymbol{k}-k_{av}\hat{\boldsymbol{z}}|^{2}/4\sigma_{k}^{2}}}{(2\pi\sigma_{k}^{2})^{\frac{3}{4}}},\label{eq:6.27}
\end{equation}
with $\epsilon=\sigma_{k}/k_{av}\ll1,$ for a state of positive helicity,
choosing the average momentum in the $z$ direction. The non-vanishing
expectations are found to be (to leading order in this small quantity
$\epsilon$)
\begin{align}
\bar{E}_{x}(x;\psi) & =\sqrt{k_{av}}\,\frac{e^{-|\boldsymbol{x}-\hat{\boldsymbol{z}}t|^{2}/4\sigma_{x}^{2}}}{(2\pi\sigma_{x}^{2})^{\frac{3}{4}}}\,\cos(k_{av}(z-t)),\nonumber \\
\bar{E}_{y}(x;\psi) & =-\sqrt{k_{av}}\,\frac{e^{-|\boldsymbol{x}-\hat{\boldsymbol{z}}t|^{2}/4\sigma_{x}^{2}}}{(2\pi\sigma_{x}^{2})^{\frac{3}{4}}}\,\sin(k_{av}(z-t)),\nonumber \\
\bar{B}_{x}(x;\psi) & =\sqrt{k_{av}}\,\frac{e^{-|\boldsymbol{x}-\hat{\boldsymbol{z}}t|^{2}/4\sigma_{x}^{2}}}{(2\pi\sigma_{x}^{2})^{\frac{3}{4}}}\,\sin(k_{av}(z-t)),\nonumber \\
\bar{B}_{y}(x;\psi) & =\sqrt{k_{av}}\,\frac{e^{-|\boldsymbol{x}-\hat{\boldsymbol{z}}t|^{2}/4\sigma_{x}^{2}}}{(2\pi\sigma_{x}^{2})^{\frac{3}{4}}}\,\cos(k_{av}(z-t)).\label{eq:6.28}
\end{align}
with the magnetic field always perpendicular to the electric field.
The length scale $\sigma_{x},$ satisfying $\sigma_{x}\sigma_{k}=1/2,$
sets the localization scale. These expressions are valid for $|t|\ll(k_{av}/\sigma_{k})\sigma_{x},$
where wavepacket spreading is negligible. We see Gaussian localization
in all directions, although the length scale is large for small $\sigma_{k}.$

At a fixed position, the electric field vector viewed facing into
the oncoming wave rotates counter-clockwise with time, as appropriate
for a positive helicity (this is called a \textit{left} circularly
polarized wave) \cite{Jackson1975}. We see
\begin{equation}
\int d^{3}x\,\frac{1}{2}\{\bar{\boldsymbol{E}}(x;\psi)^{2}+\bar{\boldsymbol{B}}(x;\psi)^{2}\}=k_{av},\label{eq:6.29}
\end{equation}
equal to the average energy and
\begin{equation}
\int d^{3}x\,\bar{\boldsymbol{E}}(x;\psi)\times\bar{\boldsymbol{B}}(x;\psi)=\boldsymbol{k}_{av}\label{eq:6.30}
\end{equation}
equal to the average momentum.

To explore the other extreme of localization in a small volume, we
consider the test state with
\begin{equation}
\Psi_{\lambda}(k)=\delta_{\lambda1}\,\frac{e^{-|\boldsymbol{k}|^{2}/4\sigma_{k}^{2}}}{(2\pi\sigma_{k}^{2})^{\frac{3}{4}}}.\label{eq:6.50}
\end{equation}
By the Heisenberg uncertainty principle, a large momentum spread,
$\sigma_{k},$ should give a small position spread, $\sigma_{x},$
(with $\sigma_{x}\sigma_{k}=1/2$) according to this measure of localization.
We use the three-dimensional vector representation of the standard
rotation, $R_{0}[\hat{\boldsymbol{k}}]$ (Eq. (\ref{eq:2.11})), to
perform the rotation of the polarization vector as in Eq. (\ref{eq:6.9}).
This gives
\begin{equation}
\boldsymbol{\epsilon}(k,+1)=-\frac{1}{\sqrt{2}}\begin{pmatrix}\cos^{2}\varphi\cos\theta+\sin^{2}\varphi-i\sin2\varphi\sin^{2}\frac{\theta}{2}\\
-\sin2\varphi\sin^{2}\frac{\theta}{2}+i(\sin^{2}\varphi\cos\theta+\cos^{2}\varphi)\\
-e^{+i\varphi}\sin\theta
\end{pmatrix}.\label{eq:6.51}
\end{equation}

We calculated the 18 combinations of six components along the $x,y$
and $z$ axes. We define (with $\{\hat{\boldsymbol{u}}_{1},\hat{\boldsymbol{u}}_{2},\hat{\boldsymbol{u}}_{3}\}=\{\hat{\boldsymbol{x}},\hat{\boldsymbol{y}},\hat{\boldsymbol{z}}\}$)
\begin{align}
\overline{E}_{i}(0,\sigma_{x}\rho_{j}\hat{\boldsymbol{u}}_{j}) & =\frac{1}{\sqrt{16\pi}}\,\frac{\sqrt{\sigma_{p}}}{(2\pi\sigma_{x}^{2})^{\frac{3}{4}}}\,e_{i}(\rho_{j})\quad\mathrm{for}\ i,j=1,2,3,\nonumber \\
\bar{B}_{i}(0,\sigma_{x}\rho_{j}\hat{\boldsymbol{u}}_{j}) & =\frac{1}{\sqrt{16\pi}}\,\frac{\sqrt{\sigma_{p}}}{(2\pi\sigma_{x}^{2})^{\frac{3}{4}}}\,b_{i}(\rho_{j})\quad\mathrm{for}\ i,j=1,2,3,\label{eq:6.52}
\end{align}
and plot the dimensionless $\boldsymbol{e}$ and $\boldsymbol{b}$
component functions. This introduces the length scale, $\sigma_{x},$
which we expect to be proportional to the widths of these functions.
Note that the $e_{i}(\rho_{j})$ and $b_{i}(\rho_{j}),$ written in
terms of the variable $\boldsymbol{\rho}=\boldsymbol{r}/\sigma_{x},$
are scale invariant, so these profiles hold for any $\sigma_{k}.$

\begin{figure}
\noindent \begin{centering}
\includegraphics[width=15cm]{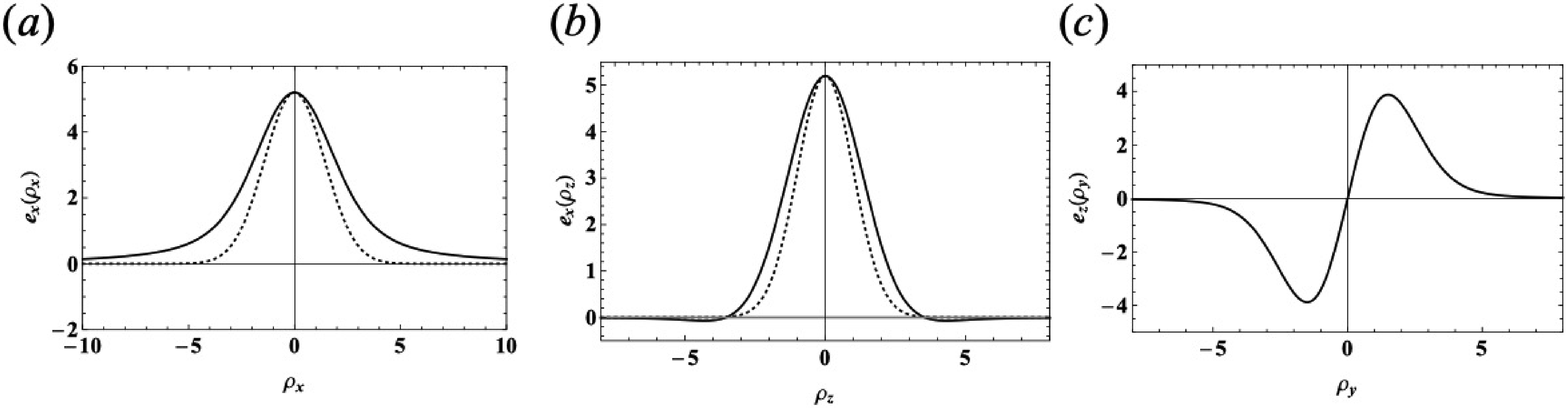}
\par\end{centering}
\caption{\label{fig:Normalized-electric-field}Normalized, dimensionless, electric
field components (a) $e_{x}(\rho_{x})$ along the $x$ axis, (b) $e_{x}(\rho_{z})$
along the $z$ axis and (c) $e_{z}(\rho_{y})$ along the $y$ axis.
In (a) $e_{x}(0)\exp(-\rho_{x}^{2}/4)$ and in (b) $e_{x}(0)\exp(-\rho_{z}^{2}/4)$
are shown for comparison.}
\end{figure}

For the electric field components, we find that $e_{y}(\rho_{x}),e_{y}(\rho_{y}),e_{z}(\rho_{x})$
and $e_{z}(\rho_{z})$ vanish identically, and that $e_{x}(\rho_{y})=e_{x}(\rho_{x})$
and $e_{y}(\rho_{z})=e_{z}(\rho_{y})/2.$ The integrals plotted are
\begin{align}
e_{x}(\rho_{x}) & =\int_{0}^{\infty}d\kappa\,\kappa^{\frac{5}{2}}\,e^{-\kappa^{2}/4}\int_{0}^{\pi}\sin\theta\,d\theta\,\{\frac{J_{1}(\kappa\rho_{x}\sin\theta/2)}{\kappa\rho_{x}\sin\theta/2}2\cos^{2}\frac{\theta}{2}-J_{2}(\kappa\rho_{x}\sin\theta/2)\cos\theta\},\nonumber \\
e_{x}(\rho_{z}) & =\int_{0}^{\infty}d\kappa\,\kappa^{\frac{5}{2}}\,e^{-\kappa^{2}/4}\,j_{0}(\kappa\rho_{z}/2),\nonumber \\
e_{z}(\rho_{y}) & =2\int_{0}^{\infty}d\kappa\,\kappa^{\frac{5}{2}}\,e^{-\kappa^{2}/4}\,j_{1}(\kappa\rho_{y}/2).\label{eq:6.53}
\end{align}
We integrated these numerically. For $e_{x}(\rho_{x}),$ it was found
possible to perform the $\kappa$ integration first, analytically
in terms of confluent hypergeometric functions, then evaluate the
$\theta$ integral numerically. These are plotted in Figure \ref{fig:Normalized-electric-field}.

\begin{figure}
\begin{centering}
\includegraphics[width=5cm]{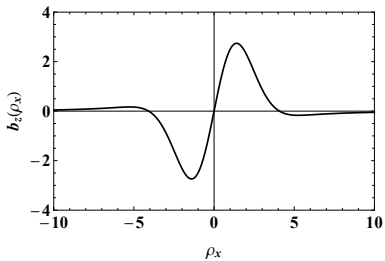}
\par\end{centering}
\caption{\label{fig:The-magnetic-field}The magnetic field component $b_{z}(\rho_{x})$
along the $x$ axis.}

\end{figure}

For the magnetic fields, we found that $b_{x}(\rho_{x}),b_{x}(\rho_{y}),b_{z}(\rho_{y})$
and $b_{z}(\rho_{z})$ vanish identically. Also $b_{y}(\rho_{x})=e_{x}(\rho_{x}),b_{y}(\rho_{y})=e_{x}(\rho_{x})$
and $b_{y}(\rho_{z})=e_{x}(\rho_{z}).$ We plot the integral
\begin{equation}
b_{z}(\rho_{x})=\int_{0}^{\infty}d\kappa\,\kappa^{\frac{5}{2}}\,e^{-\kappa^{2}/4}\int_{0}^{\pi}\sin^{2}\theta\,d\theta\,\{\frac{J_{2}(\kappa\rho_{x}\sin\theta/2)}{\kappa\rho_{x}\sin\theta/2}(\cos\theta+3)-J_{3}(\kappa\rho_{x}\sin\theta/2)\}\label{eq:6.54}
\end{equation}
in Figure \ref{fig:The-magnetic-field}. Again, the $\kappa$ integral
was performed analytically with Mathematica \cite{Mathematica2019}
before evaluating the $\theta$ integral numerically.

These field components display strong localization over a length scale
that can be made arbitrarily small by making $\sigma_{k}$ very large.
By examining the logarithms at large distances from the origin, the
asymptotic behaviour is seen to be slightly weaker than exponential
in $\rho_{j}$ for all the field components along all three axes.
We conclude that there is no lower bound on the localization length
of a photon.

Note that we found nonzero imaginary parts in the $\overline{F}^{(+)\mu\nu}(x)$
defined in Eq. (\ref{eq:6.20.1}). In our scheme, only the real parts
are physically relevant. In contrast, when calculating a photon wavefunction
(see Section \ref{sec:Comparison-with-other}) and forming the energy
density, those imaginary parts are relevant. Thus the two schemes
will give different energy densities.

We did not calculate the energy density, but it is clear that it will
have asymptotic behaviour slightly less than exponential in the coordinates.

For comparison, we find the expectations for a superposition that
is not a coherent state. It is evident that it is important to have
a contribution from the vacuum in whatever state we use to take the
expectation value. We could use the normalized state vector
\begin{equation}
|\,\beta\,\rangle=\frac{1}{\sqrt{1+|\beta|^{2}}}\{|\,0\,\rangle+\beta\,|\,\psi\,\rangle\},\label{eq:6.55}
\end{equation}
where $|\,\psi\,\rangle$ is again the one-photon state under consideration.
Then the expectations of the field strengths would be
\begin{equation}
\langle\,\beta\,|\,F^{\mu\nu}(x)\,|\,\beta\,\rangle=\frac{1}{1+|\beta|^{2}}\{\beta\,\overline{F}^{(+)\mu\nu}(x)+\beta^{*}\,\overline{F}^{(-)\mu\nu}(x)\},\label{eq:6.56}
\end{equation}
with
\begin{equation}
\overline{F}^{(+)\mu\nu}(x)=\langle\,0\,|\,F^{\mu\nu}(x)\,|\,\psi\,\rangle\quad\mathrm{and}\quad\overline{F}^{(-)\mu\nu}(x)=\langle\,\psi\,|\,F^{\mu\nu}(x)\,|\,0\,\rangle.\label{eq:6.57}
\end{equation}
\\
This measure would again show almost exponential localization, but
the dependence on the arbitrary complex parameter, $\beta,$ is undesired.
The choice $\beta=1$ gives
\begin{equation}
\langle\,\beta\,|\,F^{\mu\nu}(x)\,|\,\beta\,\rangle=\frac{1}{2}\overline{F}^{\mu\nu}(x),\label{eq:6.58}
\end{equation}
one half of our coherent state result.

\section{\label{sec:Comparison-with-other}Comparison with other work}

Sipe \cite{Sipe1995} constructed a three-vector function of spacetime
in one frame to be used as a position wavefunction for a single photon,
containing only positive energies. In our notation this is
\begin{equation}
\boldsymbol{\Psi}(x)=\int\frac{d^{3}k}{(2\pi)^{\frac{3}{2}}}e^{-ik\cdot x}\sum_{\lambda=\pm1}\boldsymbol{\epsilon}(k,\lambda)\,\sqrt{\omega}\,\Psi_{\lambda}(k).\label{eq:6.17}
\end{equation}
The intent was to make $\boldsymbol{\Psi}^{*}(x)\cdot\boldsymbol{\Psi}(x)$
act like an energy density rather than a probability density. The
integral over all space of this proposed density is
\begin{equation}
\int d^{3}x\,\boldsymbol{\Psi}^{*}(x)\cdot\boldsymbol{\Psi}(x)=\int d^{3}k\sum_{\lambda=\pm1}|\Psi_{\lambda}(k)|^{2}\,\omega=\langle\,\psi\,|\,H\,|\,\psi\,\rangle,\label{eq:6.18}
\end{equation}
the expectation of the one-photon Hamiltonian. As written, the quantities
$\boldsymbol{\Psi}(x)$ are not gauge invariant. We could construct
an alternative that is manifestly gauge invariant,
\begin{equation}
\tilde{\Psi}^{i}(x)=\int\frac{d^{3}k}{(2\pi)^{\frac{3}{2}}\omega}e^{-ik\cdot x}\sum_{\lambda=\pm1}(k^{0}\epsilon^{i}(k,\lambda)-k^{i}\epsilon^{0}(k,\lambda))\,\sqrt{\omega}\,\Psi_{\lambda}(k),\label{eq:6.19}
\end{equation}
in terms of general polarization vectors in any gauge satisfying the
Lorentz condition, Eq. (\ref{eq:6.10}). This expression reduces to
Eq. (\ref{eq:6.17}) for polarization vectors with $\epsilon^{0}(k,\lambda)=0.$

We identify these components as
\begin{equation}
\tilde{\Psi}^{i}(x)=-\sqrt{2}\,\langle\,0\,|\,F^{i0}(x)\,|\,\psi\,\rangle=-\sqrt{2}\,\langle\,0\,|\,E^{i}(x)\,|\,\psi\,\rangle,\label{eq:6.20}
\end{equation}
matrix elements of the electric field operator between the one-photon
state vector and the vacuum.

The boost transformations of these quantities could now be easily
calculated and would involve matrix elements of the magnetic fields,
$\boldsymbol{B}(x)$, not of the form of Eq. (\ref{eq:6.17}). So
just considering the electric fields does not give a complete characterization
of the system. Once this is done, the six components of
\begin{equation}
\overline{F}^{(+)\mu\nu}(x)=\langle\,0\,|\,F^{\mu\nu}(x)\,|\,\psi\,\rangle=\int\frac{d^{3}k}{(2\pi)^{\frac{3}{2}}\omega}e^{-ik\cdot x}\sum_{\lambda=\pm1}(k^{\mu}\epsilon^{\nu}(k,\lambda)-k^{\nu}\epsilon^{\mu}(k,\lambda))\,\sqrt{\omega}\,\Psi_{\lambda}(k),\label{eq:6.20.1}
\end{equation}
transform as the components of a tensor function and satisfy the zero
current Maxwell equations. (We note that $\sqrt{\omega}\,\Psi_{\lambda}(k)$
transforms under rotations and boosts with only Wigner phase factors.)
In particular, as can be seen from Eqs. (\ref{eq:2.27}) and (\ref{eq:6.11}),
the positive energy electric and magnetic field strengths rotate locally
as three-vector functions.

Up to a normalization factor, the real part of the wavefunction in
Eq. (\ref{eq:6.20.1}) is our electric field expectation value. So
we can write our expectation as
\begin{equation}
\overline{F}^{\mu\nu}(x)=\langle\,0\,|\,F^{\mu\nu}(x)\,|\,\psi\,\rangle+\langle\,\psi\,|\,F^{\mu\nu}(x)\,|\,0\,\rangle.\label{eq:6.20.1.5}
\end{equation}
From this similarity, these functions are expected to provide a useful
measure of localization.

These components are not position probability amplitudes of the form
of Eq. (\ref{eq:1.2}), since $F^{\mu\nu}(x)\,|\,0\,\rangle$ are
not localized state vectors. Neither are they expectation values.

Sipe constructed an interacting theory using these wavefunctions.

We note that Landau and Peierls \cite{Landau1930} defined a wavefunction
for the photon like Eqs. (\ref{eq:6.17}) or (\ref{eq:6.20.1}) with
the factor of $\sqrt{\omega}$ absent from the integral. The components
corresponding to $\boldsymbol{E}^{(+)}$ and $\boldsymbol{B}^{(+)},$
respectively, are $\boldsymbol{\mathcal{E}}^{(+)}$ and $\boldsymbol{\mathcal{B}}^{(+)},$
and satisfy
\begin{equation}
\int d^{3}x\,\frac{1}{2}\{\boldsymbol{\mathcal{E}}{}^{(+)}(x;\psi)^{2}+\boldsymbol{\mathcal{B}}^{(+)}(x;\psi)^{2}\}=1,\label{eq:6.20.2}
\end{equation}
like the integral of a probability density rather than an energy density.
The same considerations that we noted for Sipe's wavefunctions apply
here.

Similar to Eq. (\ref{eq:6.20.1}), the position wavefunctions of Silberstein
(and earlier, Riemann) \cite{Silberstein1907a,Silberstein1907b} and
Bia\l ynicki-Birula \cite{Bialynicki-Birula1994} are the linear combinations
\begin{equation}
\boldsymbol{F}_{\pm}(x)=\frac{1}{\sqrt{2}}\langle\,0\,|\,(\boldsymbol{E}(x)\pm i\boldsymbol{B}(x))\,|\,\psi\,\rangle.\label{eq:6.21}
\end{equation}
Their measure of localization,
\begin{equation}
\rho(x)=\boldsymbol{F}_{+}^{*}(x)\cdot\boldsymbol{F}_{+}(x)+\boldsymbol{F}_{-}^{*}(x)\cdot\boldsymbol{F}_{-}(x),\label{eq:6.22}
\end{equation}
has the dimensions of an energy density and integrates over all space
to the expectation of energy.

Other authors \cite{Saari2012,Saari2005,Keller2005,Bialynicki-Birula1998}
have investigated the bounds that the Paley-Wiener \cite{Paley1934}
theorem imposes on the large-distance behaviour of measures of photon
localization such as the ones we have considered here. The theorem
concerns integrals of the form
\begin{equation}
f(t)=\int_{0}^{\infty}d\kappa\,g(\kappa)\,e^{-i\omega(\kappa)t},\label{eq:6.31}
\end{equation}
with $\omega(\kappa)$ an everywhere positive function and where the
integrand, $g(\kappa),$ vanishes for all negative $\kappa.$ The
result is that for large $|t|,$ the falloff of the absolute value
must be slower than exponential, $\exp(-|t|/\tau),$ for some scale
$\tau>0.$

The field expectations that we calculated in Section \ref{sec:Measurement-of-localization}
are all within this bound.

Bia\l ynicki-Birula and Bia\l ynicki-Birula obtained results on the
sharp localizability of photons, where there is exactly zero probability
of detecting the photon outside a finite region. We argue that such
localization does not exist in Nature. The walls of a container are
made of atoms, and there is always a finite (albeit small) probability
that the contained particle can tunnel through those atoms. This argument
can be used against the strict definition of causality that Hegerfeldt
has found to be violated \cite{Hegerfeldt1985,Hegerfeldt1980,Hegerfeldt1974}.

They find that the electric and magnetic fields cannot simultaneously
be strictly localized. Our results in Section \ref{sec:Measurement-of-localization}
are not in contradiction with theirs, since we are not considering
strict localization and we are using coherent states, for which they
claim there is no limit on strict localizability.

\section{\label{sec:Conclusions}Conclusions}

We have proposed a measure of localization for the photon, the expectation
of the electromagnetic field strength in a coherent wavepacket state
of mean photon number unity related to a one-photon state. We considered
two test states. The first has a well-resolved momentum and one value
of helicity. The field expectations take classical forms, with Gaussian
localization. However in this case the length scale of the localization
is very large because the momentum distribution is very narrow. We
know from Fourier transform theory that a position distribution can
be wider that the mean wavelength if the momentum distribution is
narrow. The other test state has a spherically symmetric momentum
probability distribution and, again, one value of helicity. The field
expectations for this case display strong localization in all directions,
with asymptotic behaviour slightly less strong than exponential in
the field position. The localization length scale here can be made
arbitrarily small by making the momentum spread large. We conclude
that there is no lower bound on the localization length of a photon
as determined by this measure.

By characterizing a photon in terms of real electromagnetic field
strength expectations, we have provided a semi-classical description
of the photon.

We considered the work of other authors on photon localization and
found no contradiction with their results.

\bibliographystyle{vancouver}

\end{document}